\documentclass[12pt]{iopart}
\usepackage{iopams}
\usepackage{graphicx}

\expandafter\let\csname equation*\endcsname\relax
\expandafter\let\csname endequation*\endcsname\relax

\usepackage{amsfonts,amsmath,amssymb}
\usepackage{url}

\begin{document}

\title{Alane adsorption and dissociation on the Si(001) surface}

\author{R. L. Smith$^{1,2,3}$, D. R. Bowler$^{1,2,3}$}
\address{$^1$London Centre for Nanotechnology, 17-19 Gordon St, London, WC1H 0AH, U.K.}
\address{$^2$Department of Physics $\&$ Astronomy, UCL\\Gower St, London, WC1E 6BT, U.K.}
\address{$^2$Thomas Young Centre, UCL\\Gower St, London, WC1E 6BT, U.K.}
\address{$^3$International Centre for Materials Nanoarchitectonics (MANA), National Institute for Materials Science\\1-1 Namiki, Tsukuba, Ibaraki, 305-0044 Japan}

\ead{david.bowler@ucl.ac.uk}

\begin{abstract}
  We used DFT to study the energetics of the decomposition of alane,
  AlH$_3$, on the Si(001) surface, as the acceptor complement to
  PH$_3$. Alane forms a dative bond with the raised atoms of silicon
  surface dimers, via the Si atom lone pair. We calculated the
  energies of various structures along the pathway of successive
  dehydrogenation events following adsorption: AlH$_2$, AlH and Al,
  finding a gradual, significant decrease in energy. For each stage,
  we analyse the structure and bonding, and present simulated STM
  images of the lowest energy structures. Finally, we find that the
  energy of Al atoms incorporated into the surface, ejecting a Si
  atom, is comparable to Al adatoms. These findings show that Al
  incorporation is likely to be as precisely controlled as P
  incorporation, if slightly less easy to achieve.
\end{abstract}

\section{Introduction}\label{introduction}

\subsection{Background}\label{background}

Ever since the transistor was first developed in 1948, dopants have been
used to control the characteristics of semiconductor devices. Although a
relatively low dopant concentration ($\approx$10\textsuperscript{13} atoms
cm\textsuperscript{-3}) is sufficient to materially change substrate
conductivity each successive reduction of device dimensions has required
a corresponding increase in dopant concentration\cite{Dennard:1974ys}. But
concentration is ultimately limited by mutual Coulombic repulsion to
about 10\textsuperscript{20} ions cm\textsuperscript{-3}.

To ensure reliable operation a device requires a statistically
significant number (100s or 1000s) of dopant ions in its active region
such as the MOSFET channel. If there are too few charge carriers
unacceptable performance variations will arise. For example, a channel
with dimensions 50x50x10 nm\textsuperscript{3}, comparable with
present-day devices, might contain as few as 100 carriers when strongly
doped to a concentration of 5x10\textsuperscript{18}
cm\textsuperscript{-3}.

If dopant atoms can be confined to a 2-dimensional sheet with local
concentration N\textsuperscript{2D} cm\textsuperscript{-2} then an
equivalent bulk concentration N\textsuperscript{3D} =
(N\textsuperscript{2D})\textsuperscript{3/2} cm\textsuperscript{-3} is
attained. So, in the MOSFET case a sheet with N =
3x10\textsuperscript{12} cm\textsuperscript{-2} placed in the channel
conduction region would suffice. This technique requires accurate
placement of the dopant atoms and can be achieved by interrupting
substrate growth during MBE or CVD. Ultimately, downscaling will require
that dopant atoms be individually addressable, rather than manipulated
in bulk through the application of charge.

Currently attention is directed towards donor single-dopant devices, and
a FET transistor based on a single phosphorus atom on a silicon
substrate has been demonstrated\cite{Fuechsle:2012jx}. The device is fabricated by
Patterned Atomic Layer Epitaxy (PALE) with phosphine
doping\cite{Lyding:1994oc,Owen:2011vd,Shen:1995gv}. In this experiment
the 
bare Si(100) surface is 
passivated by exposure to atomic hydrogen, forming a monatomic resist.
An STM in lithography mode selectively desorbs the resist from a group
of three adjacent Si dimers. During exposure to phosphine gas at room
temperature three molecules are adsorbed at dimer end positions. Each
quickly loses an H ligand which migrates to the opposite dimer end. An
annealing phase at 350$^\circ$C causes two PH\textsubscript{2} species to
recombine with H from the gas phase and desorb. As they do so, the H
atoms of the remaining PH\textsubscript{2} progressively dissociate to
the newly-vacant sites leaving a single phosphorus atom on the surface.
This is incorporated in a subsequent annealing step, ejecting an Si in
the process.

Less progress has been seen with acceptor dopants. The variety of
devices which can be fabricated with both p and n-type dopants is much
greater than when only p-type is available. These might include p-n
junction devices such as the tunnel FET or improved n-type devices that
would benefit from an increased barrier potential around active elements
e.g. qubit devices.

Historically boron has been used as an acceptor dopant, introduced
either by ion implantation or through direct addition to molten silicon.
However, it is unsuited to precision PALE application because its
diborane precursor is known to promote H desorption from
silicon\cite{Lew:2004ud,Perrin:1989vu}.  Its small size would cause a 
delta-doped layer to be strained\cite{Sarubbi:2010lr}, causing
relatively fast diffusion within bulk Si and tending to 
smear out atomically precise dopant profiles. Aluminium, adjacent to
silicon in the periodic table, may be a better choice. Unfortunately,
the phosphine analogue alane (AlH\textsubscript{3}) is not a useful
precursor, existing in a solid crystalline form at room temperature and
decomposing at higher temperatures. However, it can be synthesized by
evaporating metallic aluminium in a molecular hydrogen stream at low
pressures\cite{Breisacher:1964rp}. Alternatively, the amine alanes are donor-acceptor
complexes known to be viable precursors in thin film deposition of Al\cite{Jones:2009cy}.
The trimethylamine complex decomposes in the gaseous phase
giving alane and the tertiary amines\cite{Gladfelter:1989kj}, and it is plausible that
this reaction would also be effective in the PALE setting.

This work is motivated by the expectation that Al will emerge as a
viable acceptor dopant for Si in the PALE fabrication process. This will
complement P donor doping, increasing the range and functionality of
molecular devices. The initial goal will be creation of Si structures
with embedded delta-doped Al layers.

A survey of all possible adsorption and subsequent dissociation modes of
the alane molecule on the Si(100) surface is attempted. Although this
might seem to imply many configuration possibilities, the actual number
($\approx$ 60) remains manageable because the H atoms are required to stay near
the initial adsorption site at each dissociation. This approach is
similar to that of Warschkow (2005) for phosphine adsorption\cite{Warschkow:2005jt}
and reflects the highly selective nature of PALE deposition. The survey
reveals the relative stability of each intermediate configuration and
the dissociation pathways that are energetically favoured.

\section{Methods}\label{methods}

\subsection{Structural survey}\label{structural-survey}

We use DFT to survey all feasible AlH\textsubscript{n} structures on
Si(100). We show a progressive increase in stability as dissociation
proceeds and characterize the more stable surface configurations using
simulated STM and electron localization plots. A full kinetic analysis
will be presented in a subsequent paper\cite{Smith:2017ym}.

\begin{figure}
\includegraphics[height=6cm]{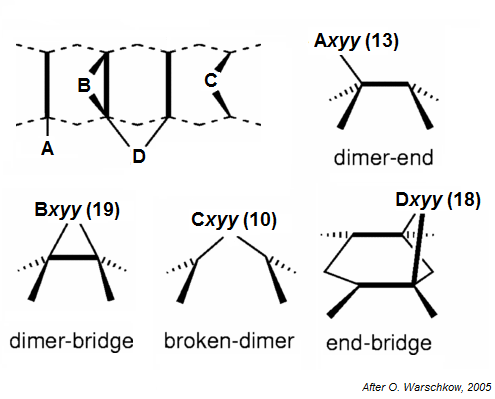} 
\caption{Perspective views of adsorption sites of
AlH\textsubscript{3} on the Si(100) surface. Adatom A binds at a
\emph{dimer-end} position of a surface dimer; B binds to two Si atoms on
the same dimer in the \emph{dimer-bridge} position, leaving the dimer
intact; \emph{broken-dimer} position C is similar to B, but breaks the
dimer and D binds to Si atoms on two adjacent dimers in the
\emph{end-bridge} position. Dissociation is modelled by removing an H
from the adatom and placing it nearby. This creates a new surface
configuration identified by appending a number \emph{xyy} where x
indicates the number of H atoms remaining bonded to Al, i.e. \emph{x=3}
represents the initial adsorption, \emph{x=0} indicates a fully
dehydrogenated Al atom. yy is an enumerator. The respective number of
identified structures appears in parentheses.\label{fig:sites}}
\end{figure}

As noted above, the initial adsorption assumes that any required
precursor reaction has already occurred and that free alane molecules
are available within bonding distance of the substrate. There are four
possible initial absorption sites as shown in Figure~\ref{fig:sites}.

\subsection{Computational details}\label{computational-details}

All calculations used density functional theory\cite{Kohn:1999qg}, as implemented
in the Vienna Ab-initio Simulation Package (VASP versions 5.3.1/5.4.1)\cite{Kresse:1996tn}
with the Perdew-Burke-Enzerhof (PBE) generalised gradient
approximation (GGA) exchange-correlation functional\cite{Perdew:1996gl}. The VASP
projector-augmented-wave\cite{Blochl:1994fq,Kresse:1999hy} (PAW) potentials for aluminium,
silicon and hydrogen were used. These potentials describe both core and
valence electrons and the files (POTCAR) were dated
4/5\textsuperscript{th} January 2001 and 15th June 2001, respectively.
We used a 400~eV energy cut-off. This value is required for proper
operation of the aluminium PAW pseudopotential.

The convergence criterion for atom forces was set to 0.02 eV/\AA\ and that
for total energy to 10\textsuperscript{-6} eV. These parameters yield
relative energies reliable to within 0.02 eV when the Brillouin zone
sampling mesh is set appropriately. For the supercell employed here,
energy values were found to converge with a 3x3x1 Monkhorst-Pack mesh\cite{Monkhorst:1976ac}.
These calculations used a quasi-Newton ionic relaxation algorithm.

\subsection{Supercell}\label{supercell}

The Si(100) surface was modelled on a slab of eight Si layers with a
c(4x2) surface cell reconstruction, separated by a 12\AA\ vacuum gap. This
surface dimension (15.36 \AA$\times$15.36 \AA) has been adopted in other studies
of this kind\cite{Brazdova:2011gx} and accommodates two dimer rows of buckled dimers
(four in each row) at approximately 18$^\circ$ to the surface plane. The
relatively large surface supports adsorption configurations spanning
adjacent dimer rows. There is less agreement over optimum cell depth,
and the chosen value is a compromise that achieves reasonable
convergence and acceptable processing times. The experimental bulk Si
lattice parameter (5.431\AA) was used and is within 1\% of the PBE
lattice constant. The bottom layer of Si atoms was left in bulk-like
positions, terminated with pairs of hydrogen atoms, and fixed.

During optimization a single AlH$x$ +(3-x)H ensemble is adsorbed on the
surface while the deepest Si and H termination layers are constrained in
fixed positions. Dissociation is modelled by progressively detaching
atoms from the Al centre and placing them elsewhere on the surface. The
energy change ${E}_{\text{AlH}_{x}}$ at each stage is calculated
by:

\begin{equation}
{E}_{\text{AlH}_{x}} = \ E_{\text{AlH}_{x} + \left( 3 - x
    \right)\text{H}} - E_{\text{Si}\left( 100 \right)} - E_{\text{AlH}_{3}}\label{eq:1}
\end{equation}

where $E_{Si(100)}$ is the energy of the clean optimized supercell,
$E_{\text{AlH}_{3}}$ the energy of an optimized alane molecule
\emph{in vacuo} and
$E_{\text{AlH}_{x} + \left( 3 - x \right)\text{H}}$ the optimized energy
of the supercell including the adsorbed AlH\textsubscript{x} and
dissociated (3-x)H species.

\subsection{Electron localization function
(ELF)}\label{electron-localization-function-elf}

ELF\cite{Becke:1990pz} is a function of the spatial coordinates which is large in
regions where electron pair density is high such as covalent bonds and
lower in regions of delocalized electronic density. It provides a useful
quantitative representation of the chemical bond in molecules and
crystals\cite{Savin:1997eu}, and is employed here to depict
$H_{0-3}Al \leftrightarrow Si(100)$ interactions. The function can be
computed from the orbitals as the definition is:

\begin{eqnarray}
\eta\left( \mathbf{r} \right) &=& \ \frac{1}{1 + \left( \frac{D}{D_{h}} \right)^{2}},\\
D &=& \frac{1}{2}\sum_{1 = 1}^{N}\left| \nabla\psi_{i} \right|^{2} -
  \frac{1}{8}\frac{\left| \nabla\rho \right|^{2}}{\rho},\\
D_{h} &=& \frac{3}{10}\left( 3\pi^{2}\right)^{\frac{2}{3}}\rho^{\frac{5}{3}},\label{eq:2}
\end{eqnarray}

with the electronic density $\rho(\mathbf{r})$ given by:
\begin{equation}
\rho = \sum_{i = 1}^{N}{\left| \psi_{I} \right|^{2},}\label{eq:3}
\end{equation}
and the sums are over the singly-occupied Kohn-Sham (or Hartree-Fock)
orbitals \(\psi_{i}\left( \mathbf{r} \right)\).
$D\left( \mathbf{r} \right)$ is the probability of finding an electron
near a reference electron of the same spin, and
$D_{h}\left( \rho\left( \mathbf{r} \right) \right)$ is the value of
$D\left( \mathbf{r} \right)$ for a homogeneous electron gas. It is
interesting to note the same dependency on kinetic energy density (the
Laplacian of the orbitals) that occurs in `meta-GGA' functionals e.g.
TPSS\cite{Tao:2003vv}. The ELF formulation inverts
$D\left( \mathbf{r} \right)$ and rescales it with respect to the HEG.
A low probability, leading to a high ELF, implies a localized electron
and vice versa. A perfectly localized orbital, such as the
H\textsubscript{2} bonding orbital, would have an ELF of 1. High ELF in
an interatomic region can be interpreted as covalent bonding, with any
asymmetries attributed to bond polarity. The HEG represents a fully
delocalized state with an ELF of 0.5. Values lower than 0.5 can
interpreted as nodal accumulations from higher order orbitals occurring
in the inter-atomic region. However, the ELF generally passes through
zero between local maxima, termed \emph{attractors}. An isosurface value of 0.8
has proven to be a useful bonding indicator in classical valence
compounds.

For the high stability configurations, we show the ELF as contour plots
in sections through the supercell. For dimer-end, dimer-bridge and
broken-dimer configurations the section is the vertical plane containing
the Al atom and the dimer, unless otherwise noted. For the other
configurations, the plane is usually horizontal or parallel to the dimer
row. The chosen isovalues are separated by an interval of 0.2, with an
additional contour in the high ELF region.

A complete set of ELF plots is available on figshare\cite{Smith:mz}.

\subsection{Simulated STM images}\label{simulated-stm-images}

Simulated STM images can show that a theoretical adsorption
configuration has an electronic structure compatible with experimental
appearance. Conversely, they can aid the identification of experimental
images. Therefore, we provide topographical (constant current) images
for the high stability configurations discovered in our survey. These
have been prepared using the Tersoff-Hamann approximation\cite{Tersoff:1985mz}as
implemented for by the bSKAN 3.3 program\cite{Hofer:2003rw}. Under this
approximation the tunnelling current is proportional to the local
density of surface states at the centre of the STM tip, whose own
electronic structure is not explicitly modelled.

We show representative simulated surface images for both positive (1.5V)
and negative (-2.0V) bias voltages. The positive value indicates current
flow into unoccupied surface states (electrons move from tip to surface)
and the negative a flow from occupied surface states (electrons move
from surface to tip).

A complete set of STM images is available on figshare\cite{Smith:mz}.

\section{Results and discussion }\label{results-and-discussion}

\subsection{Overview of the entire decomposition
pathway}\label{overview-of-the-entire-decomposition-pathway}

\begin{figure}
  \centering
\includegraphics[width=0.7\linewidth]{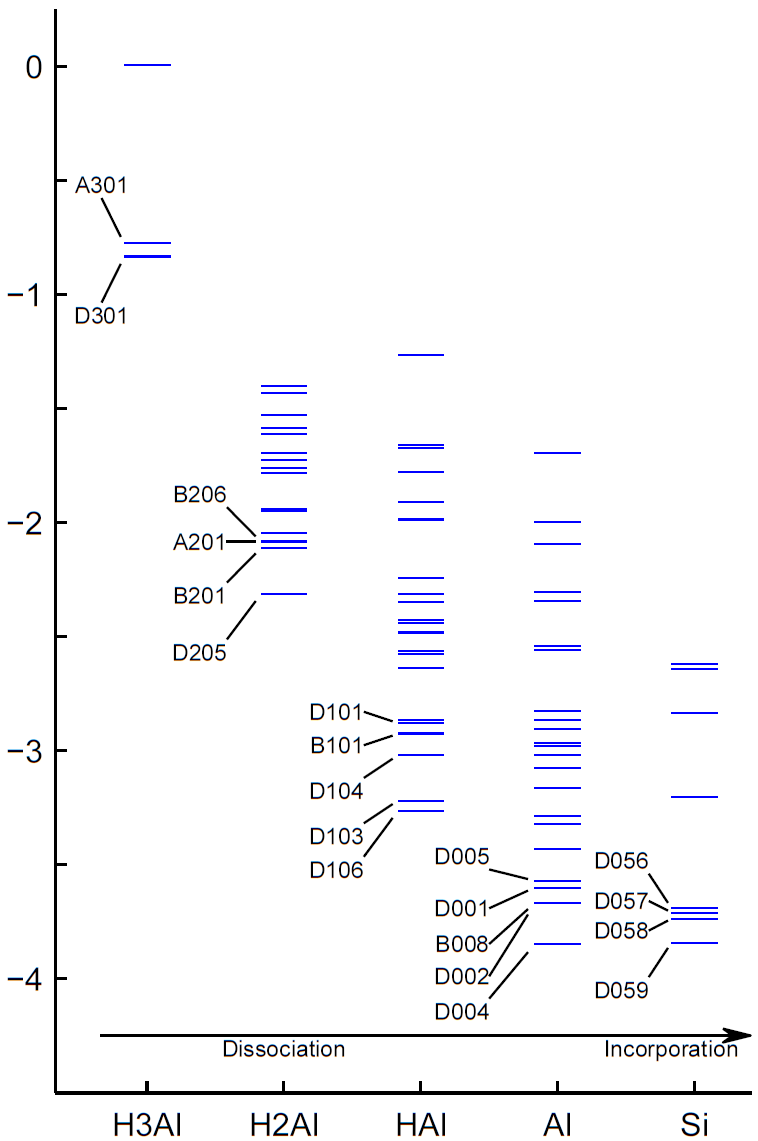}
\caption{Overview of relative energies (in eV) for alane dissociation and
  incorporation configurations considered in the
  survey. Configurations are grouped on the horizontal axis by the
  degree of dissociation. High stability (low energy) configurations
  are labelled using the numbering scheme outlined at
  Figure~\protect\ref{fig:sites}. Configuration energies are relative
  to the sum of bare surface and free alane energies. A full listing
  of the structures and relative energies is available on
  figshare\cite{Smith:mz}.\label{fig:energies}}
\end{figure}

Some 60 configurations were evaluated, showing progressive increase in
stability as dissociation proceeds. Figure~\ref{fig:energies} shows the calculated
energies as columns of bars versus the dissociation stage horizontally.
Stability of a configuration depends on the nature of the Al-Si bonding,
and the local disposition of the adsorbed H atoms. Eight incorporation
configurations are shown to demonstrate feasibility of these structures.

In each group, there are a few structures notably more stable than any
other in the same group, and a thermodynamically favoured dissociation
pathway is likely to involve these configurations. We have characterized
these \emph{high stability} structures using ELF and simulated STM
plots, and show their relative energies and bond lengths in Table~\ref{tab:energies}. Of
course, some structures may be rendered inaccessible by kinetic
considerations, and an analysis based on DFT NEB (nudged elastic band)
calculations will be the subject of another paper\cite{Smith:2017ym}.

\begin{table}
  \begin{tabular}{lllll}
& Config. & $\Delta$E (eV) & Al-Si (\AA) & Si-Si (\AA)\\
H$_3$Al: initial adsorption & D301 & -0.84 & 2.58/\emph{(4.41)} &
2.37/2.36\\
& A301 & -0.78 & 2.61 & 2.39\\
& B301 & +0.01 & \emph{(4.01)}/\emph{(4.15)} & 2.36\\
H$_2$Al: first dissociation & D205 & -2.31 & 2.55/2.62 &
2.35/2.40\\
& B201 & -2.11 & 2.45/2.81 & 2.43\\
& B206 & -2.08 & 2.48/2.54 & 2.54\\
& A201 & -2.08 & 2.49 & 2.44\\
HAl: second dissociation & D106 & -3.27 & 2.49/2.49 &
2.42/2.43\\
& D103 & -3.22 & 2.46/2.51 & 2.38/2.42\\
& D104 & -3.02 & 2.44/2.58 & 2.37/2.50\\
& B101 & -2.93 & 2.40/2.43 & 2.48\\
& A103 & -2.49 & 2.58/2.63 & 2.48/2.52\\
& C106 & -2.24 & 2.43/2.46 & \emph{(3.90)}\\
Al: third dissociation & D004 & -3.85 & 2.48/2.49 &
2.42/2.46\\
& D002 & -3.67 & 2.47/2.47 & 2.41/2.42\\
& B008 & -3.67 & 2.42/2.60/2.63 & 2.40/2.53\\
& D001 & -3.60 & 2.46/2.47 & 2.42/2.42\\
& D005 & -3.57 & 2.48/2.48 & 2.38/2.39\\
& A001 & -3.28 & 2.61 & 2.41\\
& C004 & -3.07 & 2.37/2.38 & \emph{(4.75)}\\
Si: incorporation & D059 & -3.84 & 2.42/2.42/2.44 &\\
& D058 & -3.74 & 2.39/2.39/2.44 &\\
& D057 & -3.71 & 2.39/2.39/2.45 &\\
& D056 & -3.69 & 2.40/2.40/2.47 &\\
& C050 & -3.29 & 2.40/2.41/2.45 &\\
  \end{tabular}
\caption{Calculated relative energies and bond lengths for structures
  identified in Figure~\protect\ref{fig:energies} and discussed in the
  text. In the initial adsorption and dissociation cases, the Al-Si
  column gives the length of the surface bond(s) with the adsorbed
  Al. For the incorporation cases the lengths of the two subsurface
  bonds are given, followed by the length of the Al-Si
  heterodimer. Column Si-Si gives the length of the adsorbing
  dimer(s). For comparison, dimer length on the reconstructed bare
  Si(100) surface is $\approx$ 2.36\,\AA\ in this supercell. Values in
  parentheses indicate inter-atomic distances, i.e. the absence of
  bonding.\label{tab:energies}} 
\end{table}

\subsection{The Si(100) surface}\label{the-si100-surface}

\begin{figure}
\includegraphics[height=4cm]{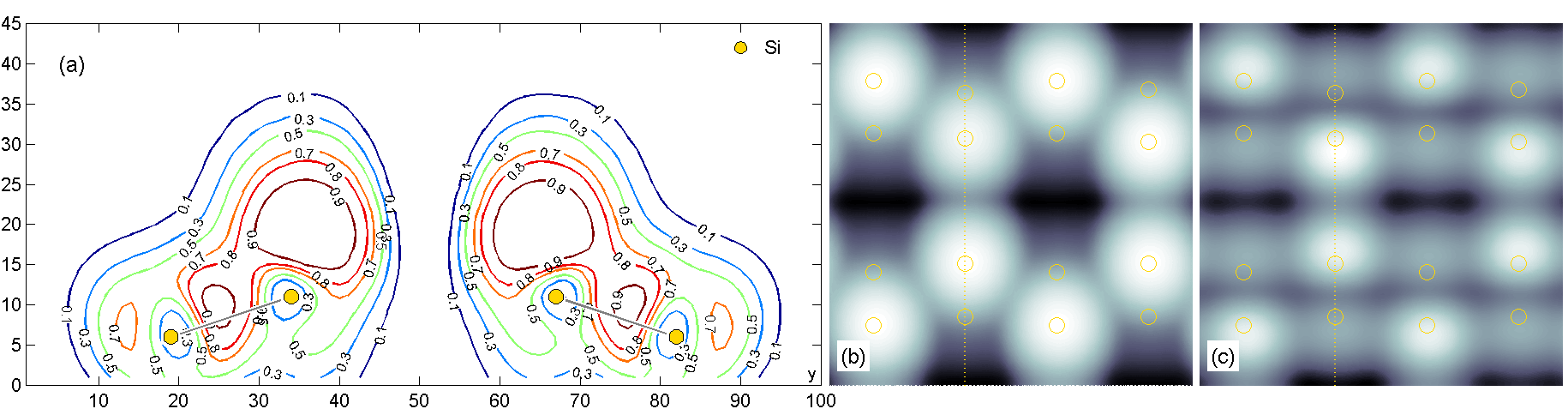}  
  \caption{ELF plot and simulated STM images for the bare
    reconstructed Si(100) surface. The ELF plot (a) is perpendicular
    to the surface in a plane containing a pair of dimers
    (indicated). The simulated STM images (b) and (c) correspond to
    tip bias voltages of -2.0V and +1.5V respectively, and the yellow
    dotted lines mark the position of the contour plane. The
    superimposed yellow circles indicate Si dimer atoms. The
    horizontal scale is 100 = 15.36 \AA.}
  \label{fig:Si001ELF}
\end{figure}

Figure~\ref{fig:Si001ELF} shows ELF and simulated STM output for the bare, reconstructed
Si(100) surface. The alternately buckled dimers are 0.2 eV more stable
than when parallel to the surface plane and are the most stable
reconstruction possible. At ambient temperatures, the dimers `flip' at a
rate greater than the STM can accommodate, and so the STM images shown
may not be observed. However, the presence of an adsorbing Al or H atom
will be sufficient to `pin' the dimer in the buckled configuration,
justifying use of the reconstruction.

The STM filled state plot shows that reconstruction eliminates one
dangling bond and concentrates electronic density at the `up' dimer end,
and dimer length is found to be 2.36~\AA\ in this supercell. The filled
state STM plot shows the DOS centred on the surface atoms. As the ELF is
determined over occupied states it might be expected to correspond with
the filled-state STM image, although no theoretical basis has been
established for this. However, the plot reveals large attractor regions
above the `up' dimer ends with ELF values exceeding 0.9, characteristic
of a non-bonding (lone) electron pair.

\subsection{Initial adsorption: H$_3$Al $\leftrightarrow$ Si(100)}\label{initial-adsorption-h3alsi100}

\begin{figure}
\includegraphics[height=3cm]{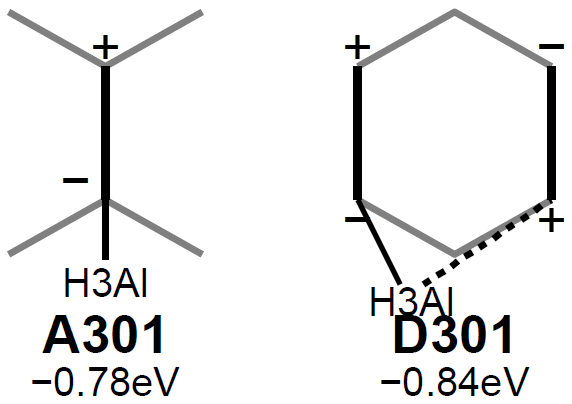}    
  \caption{Schematic representation of initial adsorption
    configurations after structural optimization, showing relative
    energies. Dimers are represented by heavy vertical lines; the
    ‘-/+‘ signs indicate the ‘up/down’ ends, respectively. All
    configuration files are available on figshare\protect\cite{Smith:mz}.}
  \label{fig:initads}
\end{figure}

\begin{figure}
\includegraphics[height=4cm]{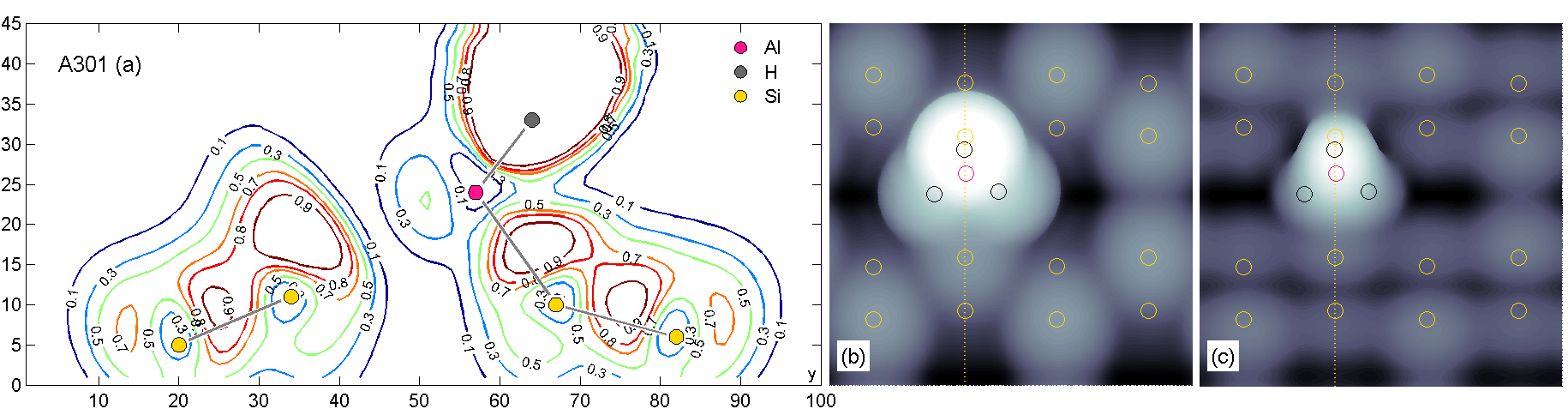}      
  \caption{ELF plot and simulated STM images for the initial dimer=end
    configuration A301. The ELF plot (a) indicates the Al-Si, Al-H and
    Si-Si bonds. The high-ELF region surrounding the H ligand shows
    the polar nature of the Al-H bond. In the STM images (b) (c), the
    Al, Si and H atom locations are superimposed and yellow dotted
    lines mark the position of the contour plane. Images (b) and (c)
    correspond to tip bias voltages of -2.0 V and +1.5V respectively.}
  \label{fig:AdsELF}
\end{figure}

Stable configurations were discovered at dimer-end, dimer-bridge and
end-dimer sites. No stable broken-dimer configuration was found, with an
H atom tending to detach and migrate to the adjacent dimer row or adopt
a central position `buried' beneath the dimer. The dimer-bridge
configuration (B301) showed a slight surface repulsion and was not
considered further. Figure~\ref{fig:initads} shows the two remaining structures and
their relative energies and Figure~\ref{fig:AdsELF} shows the ELF plot and simulated
STM images for the dimer-end configuration A301 where a bond with
stability -0.78eV was found with the up-atom and the Al atom in
pyramidal coordination. No bonding was possible with the down-atom.

Alane has six electrons in its valence shell and can accept a further
two to complete its octet. These are provided by the excess electronic
density at the surface dimer `up' end, and form a dative bond with alane
acting as a Lewis acid and the substrate as Lewis base. The Al-Si bond
length of 2.61 \AA\ obtained here can be compared with 2.08 \AA\ calculated
for the dative Al-N bond in ammonia alane\cite{Craig:1995wq}. Although both have
\emph{sp\textsuperscript{3}} hybridization the latter has greater
\emph{s} character due to the H ligands of the ammonia. This, together
with the greater electronegativity of the nitrogen atom, account for the
shorter Al-N bond. However, the ELF plots of Figures~\ref{fig:Si001ELF} and~\ref{fig:AdsELF} are
consistent with the Lewis adduct model.

The adduct model implies that the end-bridge configuration possessing
two surface bonds is unfeasible. This was confirmed by our optimization
of configuration D301 which resulted in an asymmetrical configuration
with only one bond substrate bond (see Table~\ref{tab:energies}) and the ELF plot (not
shown) confirmed the presence of just a single bond. The increased
stability (0.06 eV) compared with the dimer-end configuration is due to
a relative rotation of the alane molecule which was not explored during
the optimization of the dimer-end configuration.

These results show the initial alane adsorption modes are analogous to
those of phosphine, which bonds into the `down' atom at dimer-end sites,
but is unstable in the dimer-bridge, broken-dimer and end-bridge
configurations\cite{Warschkow:2005jt}.

\subsection{First dissociation: H$_2$Al+H $\leftrightarrow$ Si(100)}

\begin{figure}
\includegraphics[height=3cm]{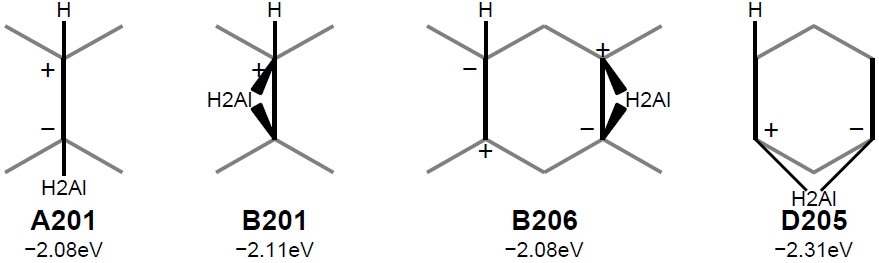}        
  \caption{Schematic representation of high stability configurations
    after the first dissociation, showing and relative energies}
  \label{fig:first_dis}
\end{figure}

\begin{figure}
  \includegraphics[height=12cm]{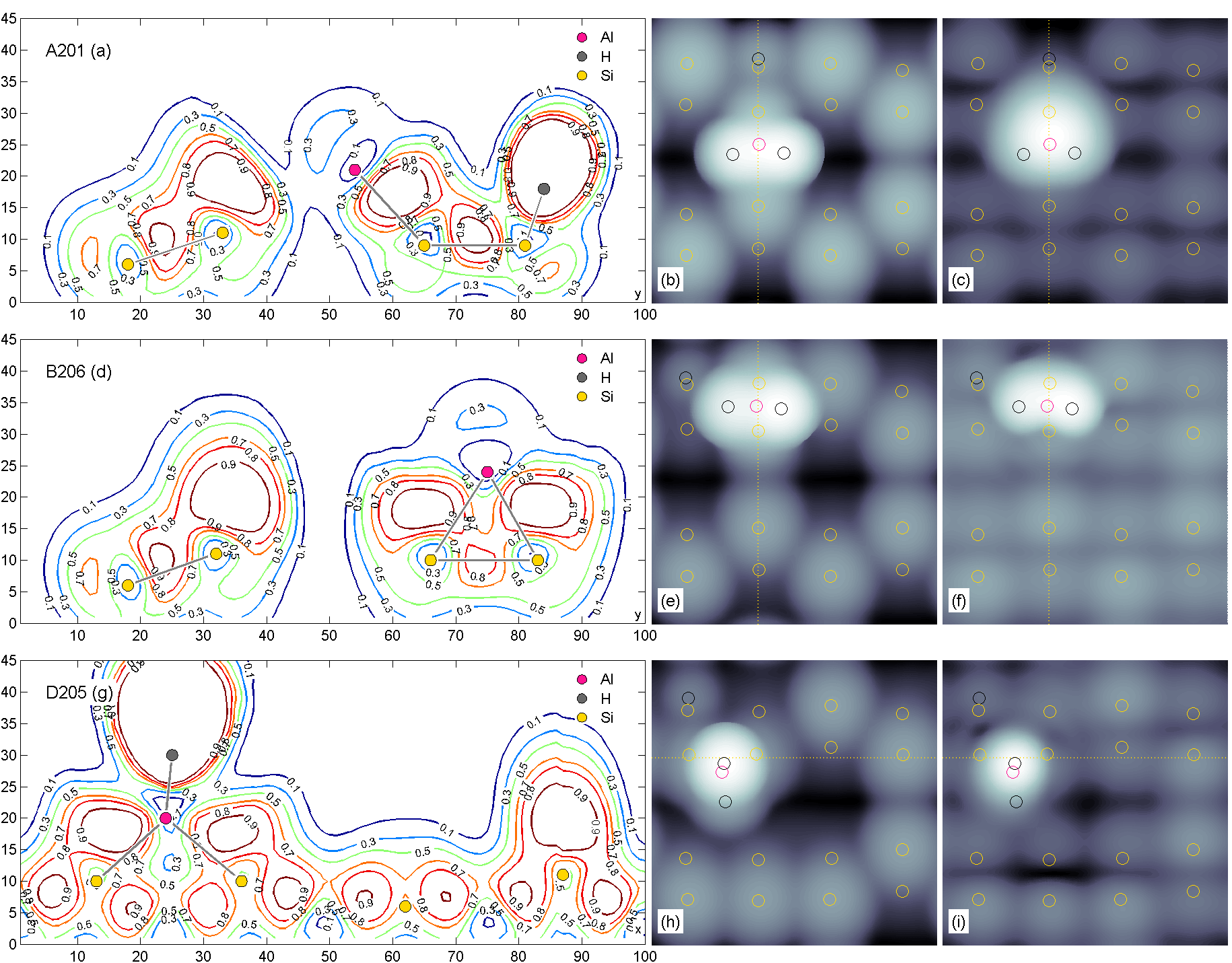}        
  \caption{ELF plots and simulated STM images for first dissociation
    configurations A201, B206 and D205. In the ELF plots (a) and (d)
    the contour map plane passes through the surface dimers
    perpendicular to the dimer row, and in (g) the plane is parallel
    to the dimer row. In the STM images, the Al, Si and H atom
    locations are superimposed and yellow dotted lines mark the
    position of the ELF contour plane. Images (b, e, h) and (c, f, i)
    correspond to tip bias voltages of -2.0 V and +1.5V respectively.}
  \label{fig:FirstELF}
\end{figure}

At this stage the high stability configurations show energies decreasing
by 1.3-1.5 eV below the initial adsorption, with end-bridge
configuration D205 the most stable, as shown in Figure~\ref{fig:first_dis}. In the absence of kinetic barriers,
these large margins suggest the initial configurations will be
relatively short-lived on the surface. The end-dimer A201 and end-bridge
D205 configurations were the most stable of their kind by margins of 0.3
and 0.5 eV respectively but two dimer-bridge configurations B201 and
B206 had similar energies, differing only in the placement of the
migrating H atom. A broken-dimer configuration appeared with the H atom
placed beneath the dimer level in an apparently three-centred bond, but
it was at least 0.5 eV less stable than the high stability group and not
considered further. Figure~\ref{fig:FirstELF} shows the ELF plots and simulated STM
images for configurations A201, B206 and D205.

The end-dimer configuration A201 has the Al atom in trigonal planar
coordination and an Al-Si bond of length of 2.49\,\AA, noticeably shorter
than that of the A301 configuration. We surmise that the ligands, now
having predominantly \emph{sp\textsuperscript{2}} hybridization, provide
improved overlap with the surface orbitals. This can be seen in the ELF
plot as an enlarged inter-nuclear region with value 0.9 or greater. The
effect of the adsorbed H on the down-dimer atom is to level the dimer,
with both atoms making 2-centre, 2-electron bonds.

In the dimer-bridge and end-bridge configurations, Al adopts a
tetrahedral configuration, although the bond angles are far from ideal.
The end-bridge configuration is the more stable by a margin of 0.2 eV. In
the dimer-bridge cases, B201 and B206, the Si-Si dimer bond lengths are
2.43\,\AA\ and 2.54\,\AA\ respectively, with both surface atoms making 2-centre,
2-electron bonds with the metal. In the end-bridge configuration D205
the dimer bond lengths are 2.35\,\AA\ and 2.40\,\AA, closer to the bare surface
value and indicating that the stability gain occurs through sharing the
adsorption stress across surface dimers.

\subsection{Second dissociation: HAl+2H $\leftrightarrow$ Si(100)}

\begin{figure}
  \includegraphics[height=3cm]{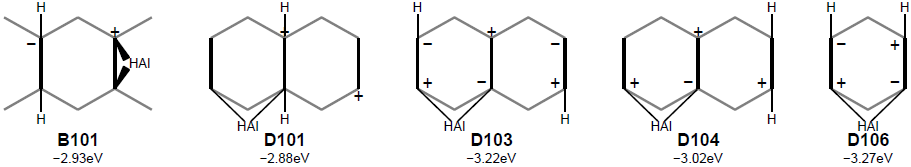}        
  \caption{Schematic representations of high stability configurations
    after the second dissociation, showing relative energies. These
    are the 5 most stable of the 28 configurations examined at this
    stage. The 10 most stable configurations were all bridged.}
  \label{fig:second_dis}
\end{figure}

\begin{figure}
  \includegraphics[height=8cm]{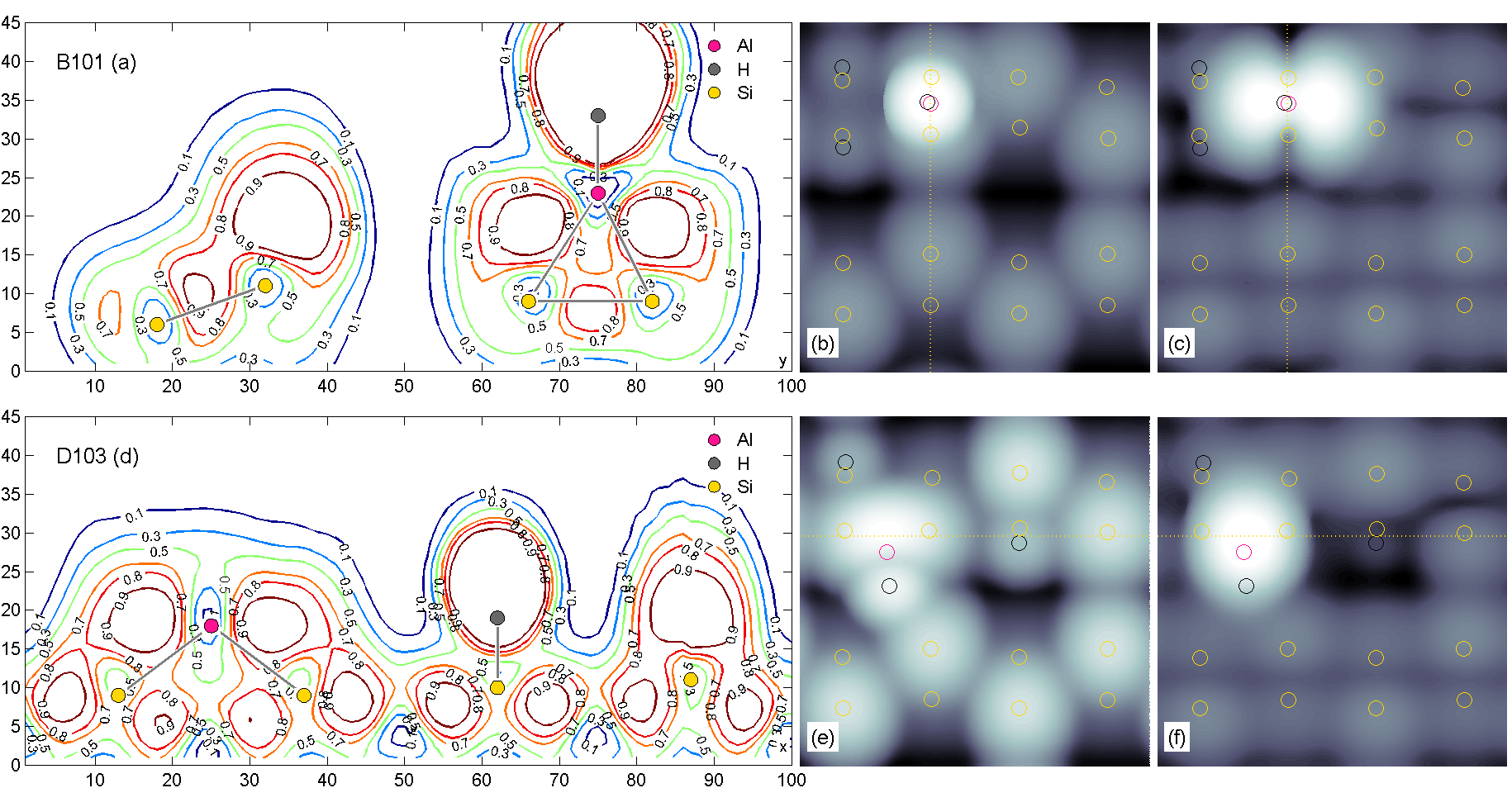}          
  \caption{ELF and simulated STM plots for second dissociation
    configurations B101 and D103. For B101 the ELF contour map (a)
    plane passes vertically through the surface dimers. For D103 (d)
    the plane is parallel to the dimer row. In the STM images, the Al,
    Si and H atom locations are superimposed and yellow dotted lines
    mark the position of the ELF contour plane. Images (b, e) and (c,
    f) correspond to tip bias voltages of -2.0 V and +1.5V
    respectively. These configurations are $\approx$0.6-0.9 eV more stable
    than at the previous stage.}
  \label{fig:SecondELF}
\end{figure}

The loss of a further H ligand increased stability by up to 0.9 eV,
depending on the surface configuration. In the absence of significant
kinetic barriers, this energy loss would prompt dissociation in the PALE
environment. The high stability configurations are all end or
dimer-bridge (see Figure~\ref{fig:second_dis}); these are the configurations likely to
appear on a pathway towards complete dissociation and incorporation. We
take D103 as representative of the end-bridge configurations and show
ELF plots and simulated STM images for B101 and D103 in Figure~\ref{fig:SecondELF}.

The ELF plots are like those of the bridged configurations of the
previous stage (see Figure~\ref{fig:FirstELF}) with the Al atom now adopting a trigonal
planar, rather than a tetrahedral coordination. In the dimer-bridge case
B101 the adsorbate bonds shorten to 2.40 \AA\ and 2.43 \AA\ compared
to 2.48 \AA\ and 2.54 \AA\ in B206, allowing the dimer bond to shorten to 2.48 \AA\ from
2.54\AA.  This improved bonding can be attributed to the increased \emph{s}
character of the adsorbate bonds feeding into the dimer bond. As before,
the sharing of surface stress in the end-bridge configuration D103 is
responsible for its additional ($\approx$0.3 eV) stability.

\begin{figure}
  \includegraphics[height=5cm]{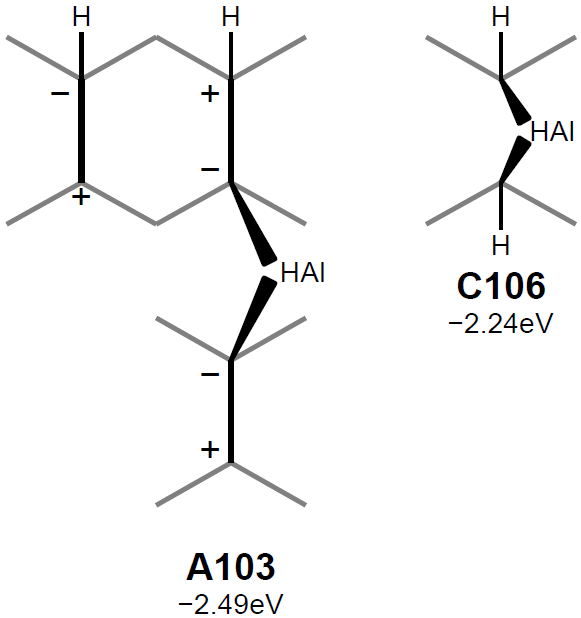}          
  \caption{Schematic representation of the most stable end and broken-
    dimer configurations after second dissociation. Structural
    optimization of dimer-end configuration A103 has moved the HAl
    fragment to a position bridging dimer rows.}
  \label{fig:second_dis2}
\end{figure}

The most stable end and broken-dimer configurations are almost 0.5 eV
less stable, and are depicted in Figure~\ref{fig:second_dis2}. Although their relative
stabilities indicate they are unlikely to participate in a dissociation
pathway they are of interest because they show the HAl fragment
preserving a trigonal planar coordination with the substrate surface.

\begin{figure}
  \includegraphics[height=4cm]{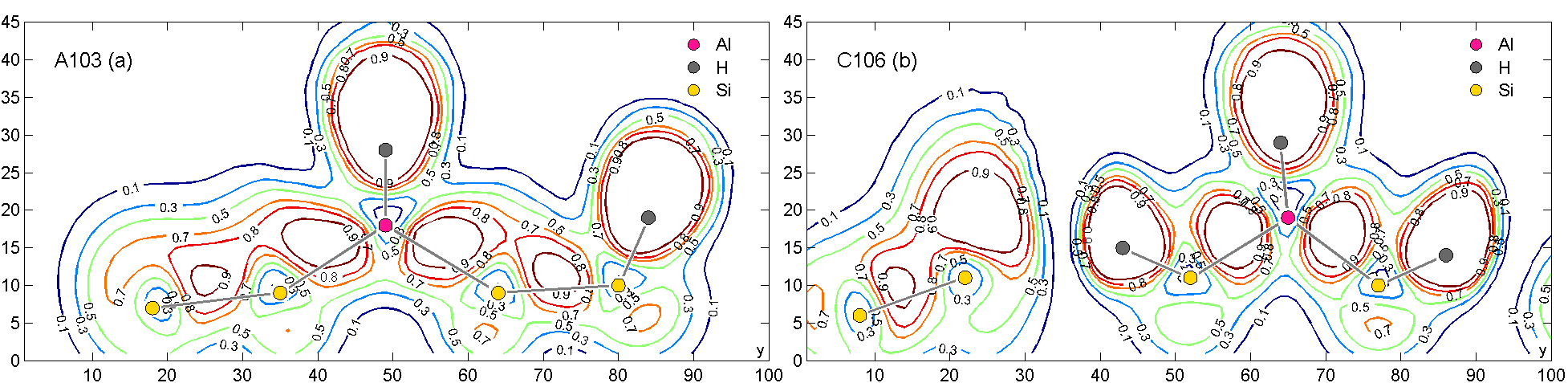}          
  \caption{ELF plots of end-dimer (a) and broken-dimer (b)
    configurations after the second dissociation. Both show the HAl
    fragment in trigonal planar coordination with the Si(100) surface,
    bridging a single dimer row (C106) or adjacent tows (A103). These
    are the most stable configurations of their type, but are $\approx$\,0.5 eV
    less stable than any bridged and end-dimer configuration at this
    stage.}
  \label{fig:SecondELF2}
\end{figure}

The corresponding ELF plots are shown in Figure~\ref{fig:SecondELF2}. Configuration A103
in Figure~\ref{fig:SecondELF2}(a) shows the Al atom located in a trigonal planar
coordination between dimer rows, bridging to an up-dimer atom in each.
Although the Si-Al-Si bond angle is a near perfect 119$^\circ$ the lack of
stability is due to the elongated adsorbate and Si-Si dimer bonds of
this configuration (Table~\ref{tab:energies}). Similar results were seen in several other
bridged-row configurations in the survey. In the broken-dimer
configuration C106 (b) the Al-Si bonds are shorter but loss of the Si-Si
dimer bond outweighs any gain in stability.

\subsection{Third dissociation: Al+3H $\leftrightarrow$ Si(100)}

\begin{figure}
  \includegraphics[height=5cm]{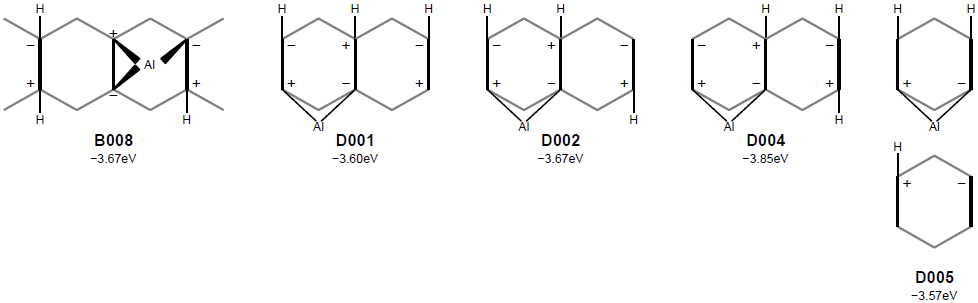}
  \caption{Schematic representations of high stability configurations
    after the third dissociation. In configuration B008 structural
    optimization has moved the Al atom from its starting dimer-bridge
    position to a mid-dimer location making three surface bonds. In
    configuration D005 a H atom has been placed on the adjacent dimer
    row, but the Al does not bridge the rows.}
\label{fig:third_dis1}
\end{figure}

27 configurations were examined; all types were represented but the
eight most stable were all the end or bridged-dimer variety. The most
stable configuration D004 gains approximately 0.6\,eV stability over its
counterpart at the previous stage; a smaller energy loss than was seen
in the first and second dissociations. Several configurations in the
survey had increased energies, reflecting the reduced coordination
possibilities available at this stage. The five most stable 
configurations span an energy range of less than 0.2\,eV and are depicted
in Figure~\ref{fig:third_dis1}.

\begin{figure}
  \includegraphics[height=8cm]{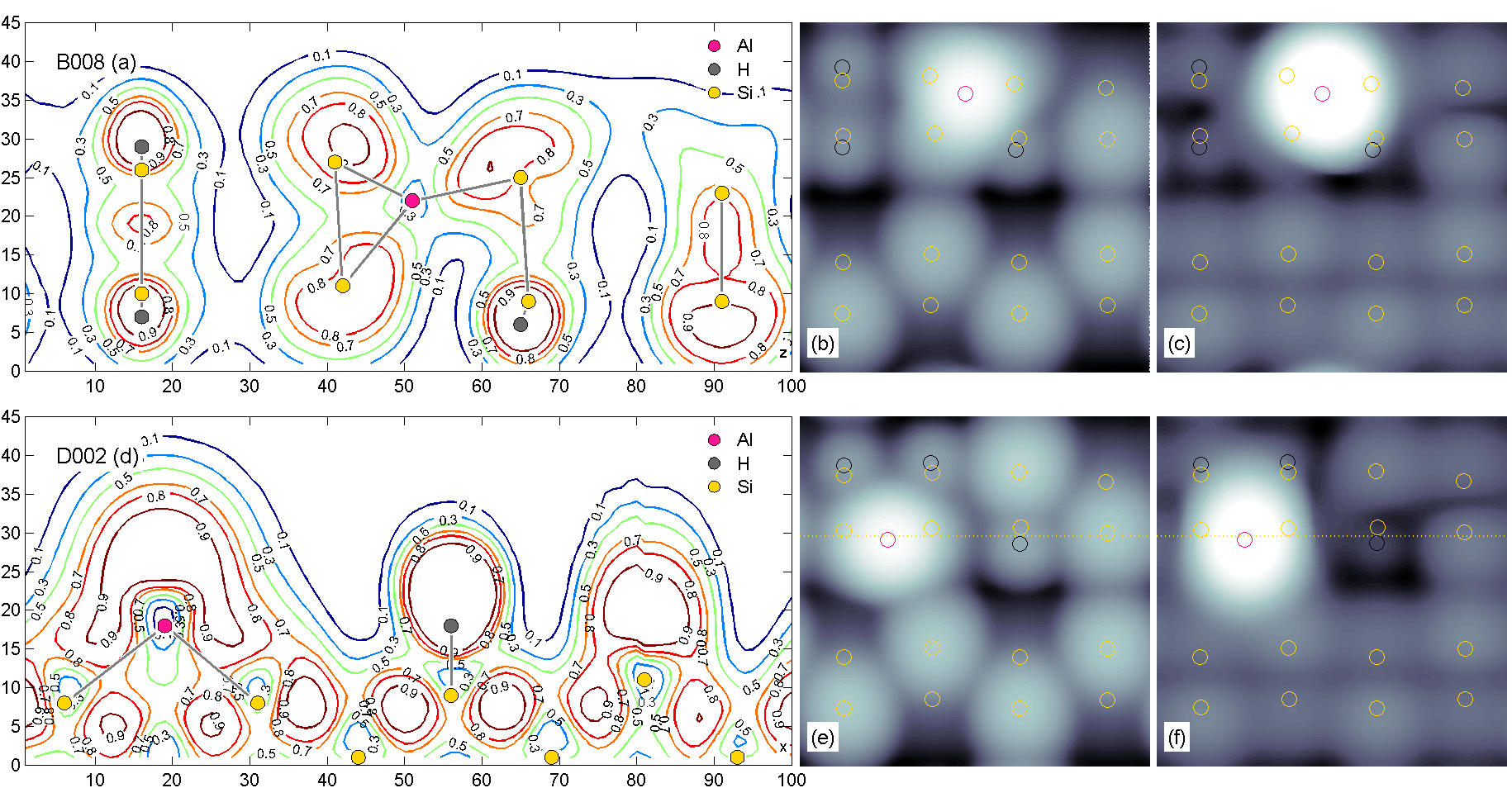}          
  \caption{ELF plots and STM images of bridged and end-dimer
    configurations after the third dissociation. The dimer-bridge ELF
    plot B008 (a) is taken in a horizontal plane (parallel to the
    surface) and shows the Al atom with three surface bonds. The
    end-bridge plot D002 (d) is taken in a vertical (perpendicular to
    the surface) plane. In the STM images, the Al, Si and H atom
    locations are superimposed and yellow dotted lines mark the
    position of the ELF contour plane. Images (b, e) and (c, f)
    correspond to tip bias voltages of -2.0\,V and +1.5\,V
    respectively. These configurations are $\approx$0.3-0.6\,eV more
    stable than at the previous stage.} 
  \label{fig:ThirdELF}
\end{figure}

After optimization, the dimer-bridge configuration B008 had the Al
adatom located between adjacent dimers, adopting a trigonal pyramidal
coordination with three surface bonds. To illustrate this the ELF plot
Figure~\ref{fig:ThirdELF}(a) is taken in the horizontal plane containing the Al atom,
above the dimers and at roughly the same elevation as nearby H atoms.
Attempts to induce a square planar Al configuration, with four surface
bonds and no surface H were unsuccessful. The bright STM images in the
unfilled state images Figures~\ref{fig:ThirdELF}(c) and (f) reflect the adsorbate's
vacant \emph{p} orbital in this coordination.

The four end-bridge configurations D001, D002, D004 and D005 are similar
in character, differing in H placement only, and we take D002 as
representative. ELF and simulated STM images for this configuration are
shown at Figure~\ref{fig:ThirdELF}(d), (e) and (f). The ELF plot Figure~\ref{fig:ThirdELF}(d) is taken
perpendicular to the surface and shows the Al adatom in trigonal planar
coordination with two surface bonds and a large hybridized lone-pair
region above.

\begin{figure}
  \includegraphics[height=3cm]{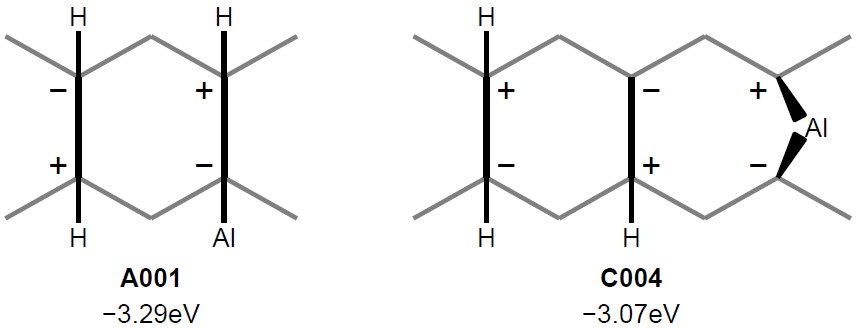}
  \caption{Schematic representation of most stable end and broken-dimer configurations after the third dissociation.}
\label{fig:third_dis2}
\end{figure}

\begin{figure}
  \includegraphics[height=4cm]{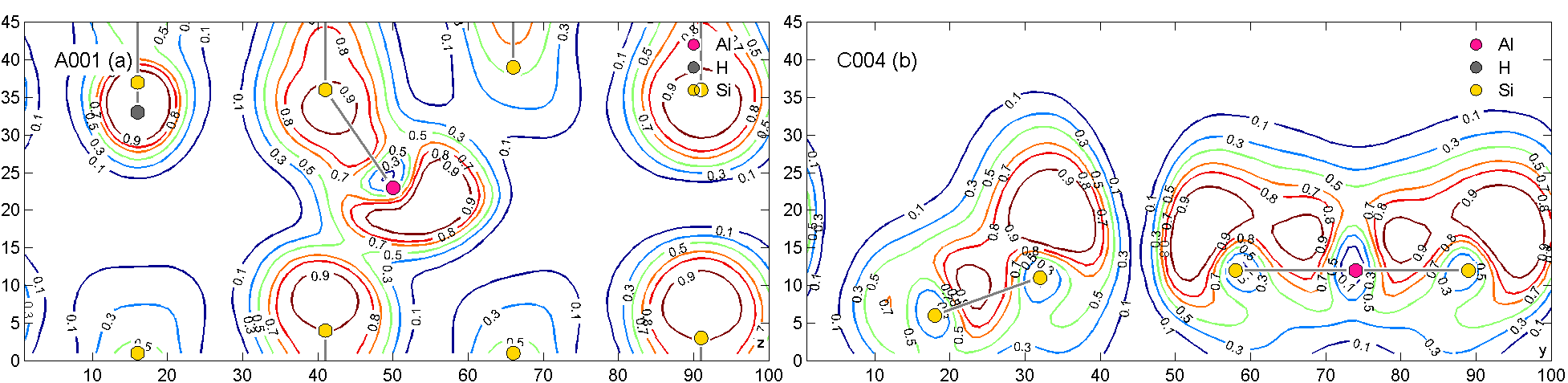}          
  \caption{ELF plots of end-dimer (a) and broken-dimer (b)
    configurations after the third and final dissociation. The plot
    for the end-dimer configuration (A001) is taken parallel to the
    surface through the midpoint of the Al-Si bond. The plot for the
    broken-dimer configuration C004 is taken in a vertical plane
    containing the dimer atoms. These are the most stable
    configurations of their type, but are respectively $\approx$0.3\,eV
    and 0.5\,eV less stable than bridged and end-dimer configurations
    at this stage.}  
  \label{fig:ThirdELF2}
\end{figure}

The most stable end (A001) and broken-dimer (C004) configurations are
shown schematically at Figure~\ref{fig:third_dis2}. Configuration A001 has Al and three H
atoms adsorbed on adjacent dimers, saturating them. The corresponding
ELF plot at Figure~\ref{fig:ThirdELF2}(a) shows the Al veering along the trench between
the dimer rows, but not bridging them as was seen for configuration A103
(Figure~\ref{fig:SecondELF2}(a) above). Here the Al-Si bond length of 2.61 \AA\ is identical
to that found in the initial adsorption case A301 with similar lengths
in the respective Si-Si dimers (Table~\ref{tab:energies}). This suggests the same dative
covalent character for the surface bond, with the unpaired Al valence
electrons arranging themselves to maximize mutual repulsion. However,
the single surface bond means that the configuration $\approx$0.3~eV less
stable than any of the bridged modes. Several other mid-trench
configurations were tried, but none proved particularly stable.

The broken-dimer configuration C004 at Figure~\ref{fig:ThirdELF2}(b) has a perfectly
linear Al coordination with predominantly \emph{sp} hybridization with
Al-Si bond lengths of $\approx$2.37\,\AA, the shortest in the survey. It is
interesting that this is a minimum energy configuration even though the
Al-Si bonds do not pass through the regions of highest ELF. However,
elimination of the surface dimer prevents any gain in overall stability,
yielding a configuration $\approx$0.5\,eV less stable than any bridged mode at
this stage.

\subsection{Incorporation}\label{incorporation}

In the PALE process the surface reaction terminates when all
unpassivated bonding sites become occupied, either by precursor
fragments or hydrogen adatoms. The dopant atoms must then be
incorporated into the surface as Si-Al heterodimers, prior to the
deposition of further Si layers. The replacement of an Si-Si dimer by
the heterodimer involves the breaking of surface bonds and requires
elevated temperatures. Successful incorporation would result in the
appearance of ejected Si atoms as surface adatoms and could be confirmed
by STM examination. After ejection from the surface the Si adatom could
reside in any one of the three bridged sites B, C or D and a systematic
survey of all heterodimer structures having three adsorbed H, an
incorporated Al and an Si adatom is beyond the present scope. Instead we
have optimized a small number of configurations of each type to
illustrate the energetics of Al incorporation.

\begin{figure}
  \includegraphics[height=3cm]{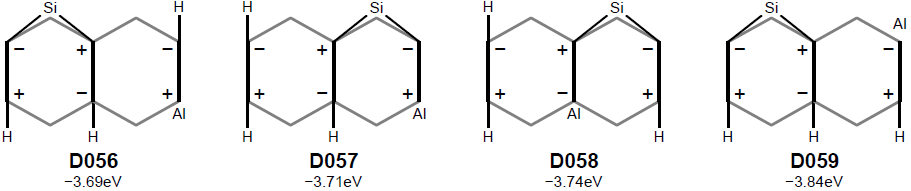}
  \caption{Schematic representation of high stability configurations
    after incorporation.}
\label{fig:incorp}
\end{figure}

We examined eight incorporation configurations. Each has an ejected Si
adatom with two surface bonds and an incorporated Al forming a Si-Al
heterodimer. Their relative energies appear in Table~\ref{tab:energies} and are
represented graphically in the rightmost column of Figure~\ref{fig:energies}. The four
configurations with the greatest stability were of the end-bridge
variety and are shown schematically at Figure~\ref{fig:incorp}. They differ only in
the placement of H atoms and fall within a 0.15~eV energy span. The most
stable (D059, -3.84 eV) has almost the same stability ($\Delta$E = 0.004 eV) as
configuration D004 at the final stage of dissociation. This margin is
less than DFT accuracy and would result in a theoretical 50\%
incorporation assuming both states were equally stable and both
kinetically accessible.

\begin{figure}
  \includegraphics[height=4cm]{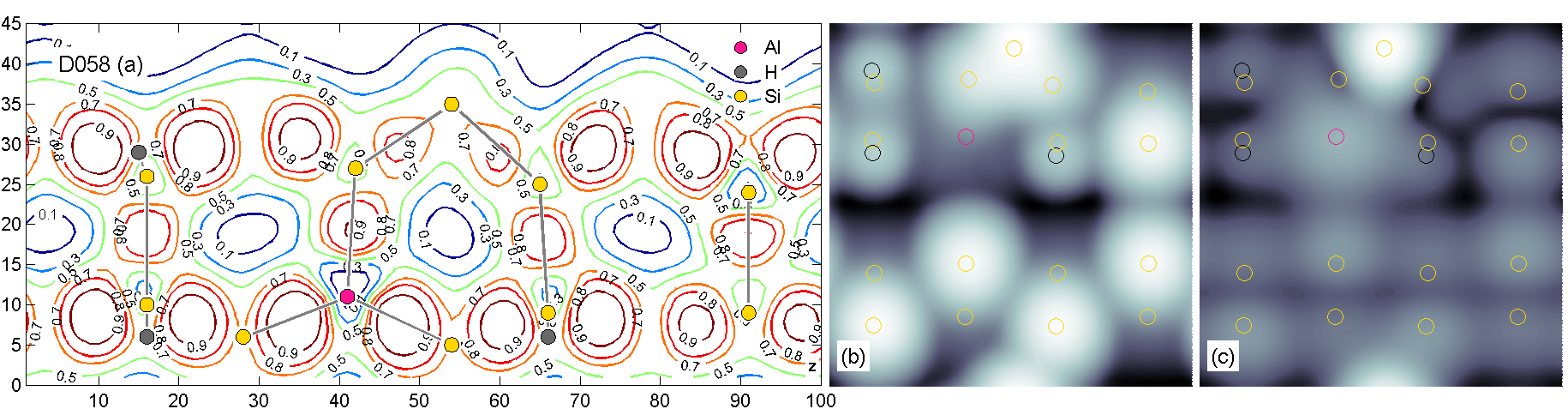}          
  \caption{ELF and simulated STM images for incorporation
    configuration D058. The ELF plot is taken in the horizontal plane
    containing the Al atom. The STM images, have the locations of the
    Al, Si and H atom locations superimposed. Images (b) and (c)
    correspond to tip bias voltages of -2.0 V and +1.5V respectively.}   
  \label{fig:IncELF}
\end{figure}

We take configuration D058 as representative and show ELF and simulated
STM plots at Figure~\ref{fig:IncELF}. Although the Al atom replaced an `up' Si it
becomes the `down' atom after incorporation. It has a pyramidal
coordination with two subsurface bonds of length 2.39~\AA\ and the
heterodimer of 2.44~\AA. The adjacent dimers are levelled. The ELF plot
Figure~\ref{fig:IncELF}(a) confirms the covalent character of these bonds. The
filled-state STM image Figure~\ref{fig:IncELF}(b) shows the absence of a dangling
bond.

\section{Conclusion}\label{conclusion}

We have used DFT to study the structure and energetics of the AlH$_x$
species which come from the adsorption and dissociation of AlH$_3$ on the
Si(100) surface, also considering several incorporation scenarios. We
find a progressive, though declining, gain in stability as the
dissociation and incorporation proceeds. The initial surface bond is
dative and tetrahedral with the adsorbate fragment adopting trigonal
geometries as dissociation proceeds. At each stage, we have identified
high stability structures likely to occur on any dissociation pathway,
and find that dimer bridging dominates. We have characterized each
structure using ELF plots and simulated STM images to aid experiment.

The energetics indicate that decomposition should be as easy to achieve
as that of PH$_3$ on Si(001); in a forthcoming paper\cite{Smith:2017ym}, we will
discuss the kinetic barriers between the dissociation fragments
described here. Overall, the energetics of the incorporated Al are close
to those of the Al adatom, in contrast to P which continues to stabilise
with incorporation. Nevertheless, there is good reason to expect at
least 50\% incorporation from Al adatoms, which may be further aided by
kinetic effects.

The ability to incorporate acceptor dopants as well as donors in Si(001)
with atomic precision will significantly advance the capabilities of
patterned ALE. It opens the possibility of p-n junctions fabricated with
atomic precision, as well as local control of the electrostatic
potential using both positive and negative dopant ions. We keenly
anticipate experimental measurements of these structures as a first
realisation of this.

\section{Acknowledgements}\label{acknowledgements}

The authors acknowledge useful discussions with James Owen and John
Randall of Zyvex Labs.

The authors acknowledge the use of the UCL Grace High Performance
Computing Facility (Grace@UCL), and associated support services, in the
completion of this work.

\bibliographystyle{jpcm}
\bibliography{biblio}

\end{document}